\documentclass[usenatbib, a4paper]{mn2e}
\usepackage{times}
\usepackage{txfonts}
\usepackage{epsfig}
\usepackage{graphicx}
\usepackage{natbib}
\usepackage{textcomp}
\usepackage[margin=0.5in, top=1in, bottom=1in, a4paper, pdftex, dvips]{geometry}

\title[Solar cycle variations in large separations]
{Solar cycle variations of large frequency separations of acoustic
modes: Implications for asteroseismology}
\author[A.-M. Broomhall et al.]{A.-M.
Broomhall$^1$\thanks{amb@bison.ph.bham.ac.uk}, W.~J. Chaplin$^1$, Y.
Elsworth$^1$, R. New$^2$\\ $^1$School of Physics and Astronomy,
University of Birmingham, Edgbaston, Birmingham B15 2TT\\
$^2$Faculty of Arts, Computing, Engineering and Sciences, Shuffled
Hallam University, Sheffield S1 1WB}

\begin{document}
\maketitle \begin{abstract}We have studied solar cycle changes in
the large frequency separations that can be observed in Birmingham
Solar Oscillations Network (BiSON) data. The large frequency
separation is often one of the first outputs from asteroseismic
studies because it can help constrain stellar properties like mass
and radius. We have used three methods for estimating the large
separations: use of individual p-mode frequencies, computation of
the autocorrelation of frequency-power spectra, and computation of
the power spectrum of the power spectrum. The values of the large
separations obtained by the different methods are offset from each
other and have differing sensitivities to the realization noise. A
simple model was used to predict solar cycle variations in the large
separations, indicating that the variations are due to the
well-known solar cycle changes to mode frequency. However, this
model is only valid over a restricted frequency range. We discuss
the implications of these results for asteroseismology.
\end{abstract}


\section{Introduction}

Helioseismic and asteroseismic frequency-power spectra contain a
rich array of oscillation peaks. The separation in frequency between
acoustic (p) modes with the same harmonic degree ($l$) and
consecutive radial orders ($n$) is known as the large frequency
separation, $\Delta\nu$. Insights into stellar structure and
evolution can be obtained by determining $\Delta\nu$ because it can
help constrain stellar properties, such as mass, radius, and log
$g$, with high precision \citep[e.g.][]{Kallinger2009, Stello2008,
Stello2009a, Miglio2009}. Therefore, one of the first aims when
analysing any new set of asteroseismic data of a solar-type star is
to determine the $\Delta\nu$ of the observable low-$l$ modes.

The frequencies of high order p modes are expected to follow
approximately the asymptotic relation \citep{Tassoul1980} and so
$\Delta\nu$ may be regarded as approximately constant. However,
$\Delta\nu$ is dependent on both frequency and $l$ and the
$\Delta\nu$ often quoted in literature is that observed for $l=0$
modes.

In this paper we describe three methods of determining the large
separation (see Section \ref{section[method]}). The most intuitive
way of determining $\Delta\nu$ is to use the individual mode
frequencies (see Section \ref{section[method fit frequencies]}).
However, sometimes the signal-to-noise ratio of the oscillations is
insufficient to allow the robust extraction of individual
oscillation frequencies. This can be the case for asteroseismic data
and can be true for solar data, for example, when the duty cycle is
low. It is therefore convenient to be able to obtain $\Delta\nu$
without determining the individual mode frequencies and so we also
obtain $\Delta\nu$ using the autocorrelation of frequency-power
spectra (Section \ref{section[method autocorrelation]}) and using
the power spectrum of the power spectrum (Section
\ref{section[method PSPS]}).

It is well known that the Sun's p-mode frequencies vary throughout
the 11-yr solar activity cycle with frequencies being at their
largest when solar activity is at its maximum
\citep[e.g.][]{Woodard1985, Palle1989, Elsworth1990, Jimenez2003,
Chaplin2007, Jimenez2007}. By examining the changes in the observed
p-mode frequencies throughout the solar cycle, we can learn about
solar-cycle-related processes that occur beneath the Sun's surface.
In Section \ref{section[results]} we show that solar cycle
variations in $\Delta\nu$ can be observed in Birmingham Solar
Oscillations Network (BiSON) data. This could have important
consequences for asteroseismic studies, which use $\Delta\nu$ to
determine fundamental stellar properties. We also show that a simple
model can be used to predict the observed variation in $\Delta\nu$
(Section \ref{section[model]}).

In Section \ref{section[comparison]} we discuss the impact of
observational choices on the obtained $\Delta\nu$. For example,
\citet{Kholikov2008} determine the solar $\Delta\nu$ using GONG and
MDI data and some intriguing differences are found between their
results and the $\Delta\nu$ obtained here. A discussion of the
implications of our results, for both astero- and helioseismology,
is provided in Section \ref{section[discussion]}.

\section{Determining the large separations}\label{section[method]}

BiSON has now been collecting data for over 30\,yr. The quality of
the early data, however, is poor compared to more recent data
because of poor time coverage. Here, we have analyzed the mode
frequencies observed by BiSON during the last solar cycle i.e. from
1996 April 11 to 2010 October 8. The precision with which p-mode
frequencies can be determined is directly related to the length of
time series under consideration. Consequently, p-mode frequencies
are often determined from time series whose lengths are of the order
of years. However, a compromise must be made here regarding the
appropriate length of time series for study so that changes during
the solar cycle can be resolved. The observations made by BiSON were
divided into 182.5-d-long independent subsets. Over the total
observation period the fractional duty cycles, or ``fills'', of the
182.5-d subsets ranged from 0.72 to 0.88.

Each method of determining the large separation must be applied over
a certain range in frequency. Initially we considered the range
$2500\le\nu\le\,\rm3700$\,\textmu Hz. At frequencies lower than
$2500$\,\textmu Hz there is very little change in frequency over the
course of the solar cycle and so it is reasonable to assume that the
large separation is also relatively constant. Above $3700$\,\textmu
Hz the relationship between frequency shift and activity changes, as
the magnitude of the solar cycle frequency shift decreases
\citep{Libbrecht1990, Chaplin1998}. Furthermore, above
$\sim4100$\,\textmu Hz it appears that modes experience a decrease
in frequency as activity increases \citep{Ronan1994, Chaplin1998}.
\citeauthor{Kholikov2008} use the range $2300\le\nu\le4300$\,\textmu
Hz and so in Section \ref{section[comparison]} we have repeated our
analysis over this frequency range. We now describe in turn each of
the three methods of determining $\Delta\nu$.

\subsection{The fitted method}\label{section[method fit
frequencies]}

\begin{figure*}
  \centering
  \includegraphics[width=0.3\textwidth, clip]{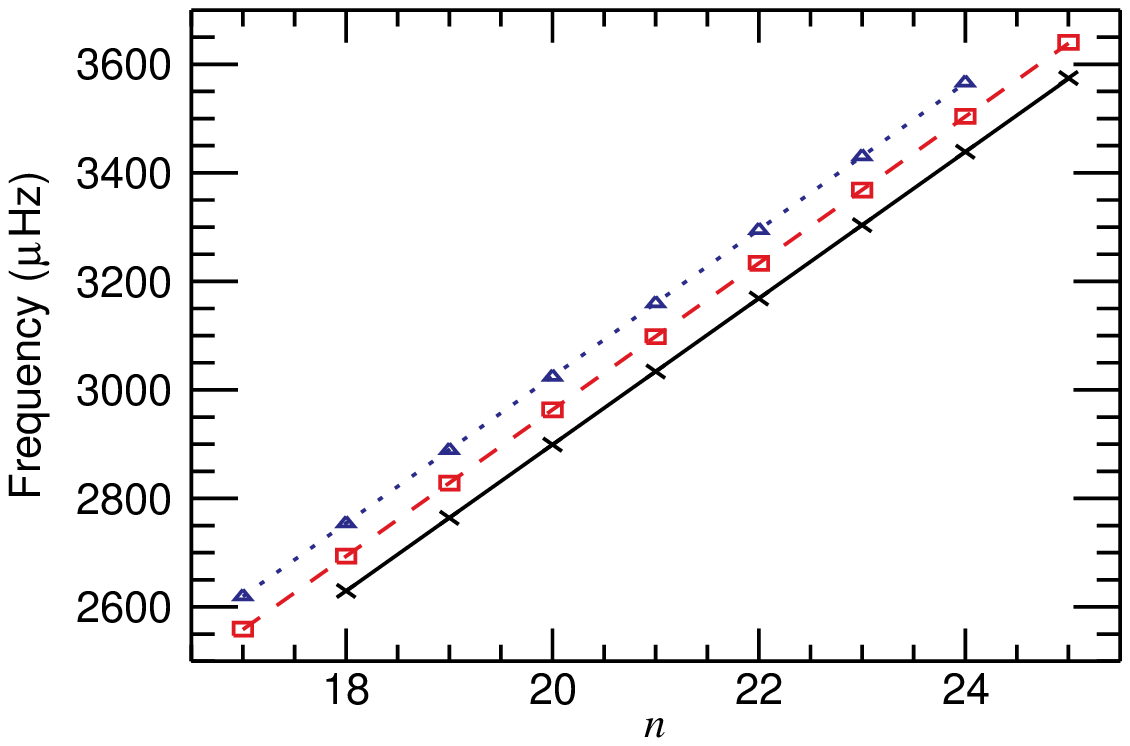}
  \includegraphics[width=0.3\textwidth, clip]{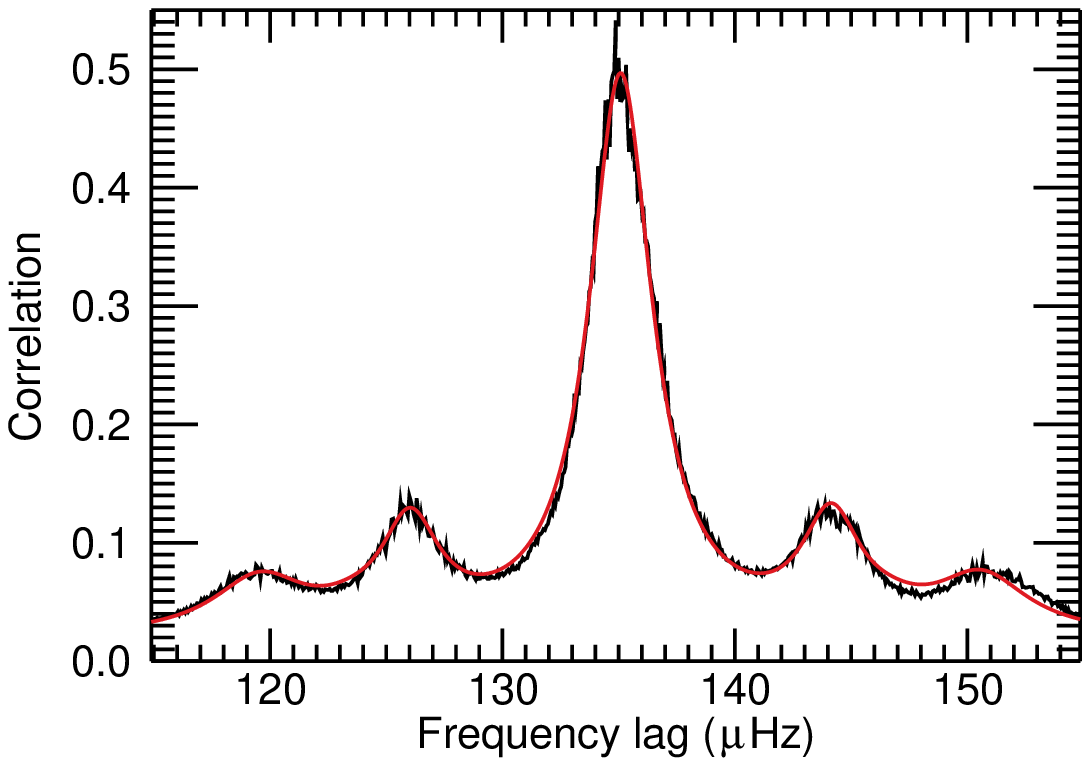}
  \includegraphics[width=0.3\textwidth, clip]{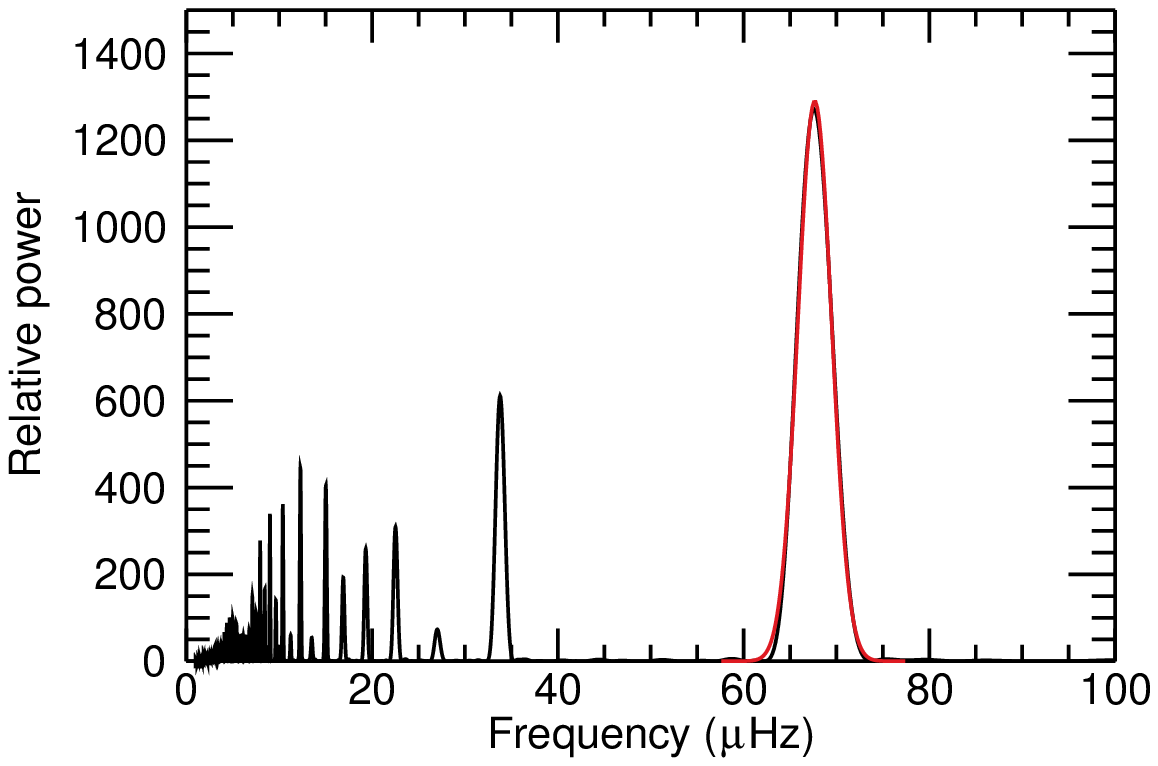}\\
  \caption{Examples of the different methods of determining
  $\Delta\nu$. Left-hand panel: Fitted method. The different symbols represent
  the fitted mode frequencies for different $l$ (black crosses
  represent $l=0$, red squares represent $l=1$, and blue triangles
  represent $l=2$). The errors
  associated with the fitted frequencies are smaller than the
  symbols. For example, the mean frequency error for the plotted graph was
  $0.12$\,\textmu Hz. The different lines show the linear fits to the observed
  frequencies (black solid line for $l=0$, red dashed line for
  $l=1$,
  and blue dotted line for $l=2$).
  Middle panel: Autocorrelation method.
  The black line shows the
  autocorrelation and the red line shows the fitted model. Right-hand panel:
  PSPS method. The black line shows the
  power spectrum of the power spectrum and the red line shows the fitted model.
  }\label{figure[examples]}
\end{figure*}

The large separations can be obtained from the individual mode
frequencies, $\nu_{l,n}$. Estimates of the mode frequencies were
extracted from each 182.5-d subset by fitting a modified Lorentzian
model to the data using a standard likelihood maximization method
\citep{Fletcher2009}. The large separations are then given by
\begin{equation}\label{equation[deltanu]}
    \Delta\nu_{l,n}=\nu_{l,n+1}-\nu_{l,n}.
\end{equation}
However, determination of the average $\Delta\nu$ using the
individual mode frequencies is complicated by the fact that
$\Delta\nu_{l,n}$ is correlated to $\Delta\nu_{l, n+1}$. Therefore a
linear function was fitted between $n$ and mode frequency,
$\nu_{l,n}$, and the gradient of this linear fit giving the average
large frequency separation, $\Delta\nu$. The linear fit was weighted
by the formal errors associated with the fitted frequencies. The fit
was performed separately for each $l$ and the mean of the gradients
was determined. The left-hand panel of Fig. \ref{figure[examples]}
shows that a linear fit represents the data well. For the remainder
of this paper we refer to this method of determining the large
separations as the ``fitted method''. Although modes with $l$ as
high as 5 can be observed in Sun-as-a-star data these modes are
unlikely to contribute significantly to the observed results as
their amplitudes are significantly smaller than modes with $l\le2$.
Therefore, when using the individual mode frequencies to determine
$\Delta\nu$ we have used information from modes with $l\le2$ only.
We have estimated the average $\Delta\nu$ for all $l\le2$ and we
have examined $\Delta\nu$ for each $l$ individually.

Monte Carlo simulations were used to determine whether the
uncertainties returned by the linear fit reflect the true
uncertainties associated with this method of determining the large
separation. 400 artificial BiSON-like 182.5-d time series were
generated \citep{Chaplin2006}. The mode frequencies used to generate
the data were shifted in the manner expected due to the solar cycle
\citep[see][and Section \ref{section[model]} of this
paper]{Broomhall2009a}. Although the magnitude of the shift placed
on each individual mode was dependent on mode frequency and $l$, the
size of the frequency shifts were based on an $l=0$ mode at
$3000$\,\textmu Hz being shifted by $0.5$\,\textmu Hz. Simulations
were also run using a smaller frequency shift but the results were
found to be relatively insensitive to the size of the input shift.
The ratio of the standard deviation of $\Delta\nu$ obtained from the
simulations and the mean uncertainty on $\Delta\nu$ indicates by how
much the observed uncertainties need to be scaled to reflect the
true errors. The scaling factors were dependent on both the
frequency range used to determine $\Delta\nu$ and which $l$ were
used. The simulated data were given various BiSON-like window
functions whose fills represented the range of fills observed here
i.e. 0.72-0.88. Although dependent on fill, over the range
$0.72-0.88$ the scaling factors were approximately constant.
Therefore the scaling factor used to determine the size of the
errors was an average of the scaling factors obtained for the
different simulated fills. When calculating $\Delta\nu$ over
$2500\le \nu_{l,n}\le 3700$\,\textmu Hz and using $0\le l\le 2$ the
scaling factor was 1.5 i.e. the true uncertainties associated with
the fitted method were 1.5 times larger than the uncertainties
implied by the fit.

\subsection{The autocorrelation method}\label{section[method autocorrelation]}

In the past cross-correlations of frequency-power spectra were used
to estimate the mean shifts in the frequencies of solar p modes.
This approach is particularly advantageous when the data have a low
S/N, which makes individual mode fitting difficult, because all of
the frequency data in the range under consideration are used
\citep{Regulo1994}. This method is also applied in asteroseismic
studies. For example, \citet{Campante2010} describe an automated
method of fitting the autocovariance of frequency-power spectra of
asteroseismic data in order to obtain, amongst other parameters, the
large frequency separation.

For different frequency lags, $\delta\nu$, the autocorrelation of a
frequency-power spectrum was determined over a given frequency
range. Let $P(\nu)$ be the frequency-power spectrum, $\nu_1$ the
lower bound of the frequency range under consideration (i.e.
$\nu_1=2500$\,\textmu Hz) and $\nu_2$ the upper limit on the
frequency range (i.e. $\nu_2=3700$\,\textmu Hz). The
autocorrelation, $C_{\scriptsize{A}}$, is then given by
\begin{equation}
C_{\scriptsize{A}}=\frac{\langle P[\nu_1:\nu_2]\cdot
P[\nu_1+\delta\nu:\nu_2+\delta\nu]\rangle}{\sqrt{\langle
P[\nu_1:\nu_2]^2\rangle\langle
P[\nu_1+\delta\nu:\nu_2+\delta\nu]^2\rangle}}.
\end{equation}

When cross-correlations were first used to determine solar-cycle
frequency shifts a second-order polynomial was fitted to the log of
the cross-correlation function over $\pm5$\,\textmu Hz around a lag
of $0$\,\textmu Hz \citep{Regulo1994, JimenezReyes1998}. However,
\citet{JimenezReyes2001} showed that more accurate results could be
obtained by fitting symmetric Lorentzians to the main peak and peaks
at the diurnal frequencies over a range of $\pm20$\,\textmu Hz. This
approach was based on the assumption that the mode peaks themselves
are Lorentzian in shape but neglects mode asymmetries. Fitting the
main peak and the diurnal sidebands works well when the duty cycle
of the data is low. However, as the fill increases the amplitude of
the sidebands decreases and other features become visible in the
cross-correlation spectrum. \citet{Chaplin2007} fitted a function
based on 7 Lorentzians: 1 for the central peak, 2 for the diurnal
frequencies, and 2 each for the $l=0,2$ and $l=1,3$ overlapping
pairs.

A structure containing the same combination of Lorentzian peaks is
observed in the autocorrelation at multiples of the large separation
(see, for example, the middle panel of Fig. \ref{figure[examples]})
and so by fitting the autocorrelation function of the
frequency-power spectrum centred on a lag of $\sim135$\,\textmu Hz
we can determine the average large separation, $\Delta\nu$. We have
fitted the peaks on both the positive and negative sides of zero lag
and then taken the average of the fitted results. The fitted
function, $M(\delta\nu)$ took the form
\begin{equation}\label{equation[model]}
M(\delta\nu)=\sum_{k=-3}^{3}\frac{\beta_kA
\left[\left(\Gamma+\delta\Gamma_{|k|}\right)/2\right]^2}
{\left[\Delta\nu-\delta\nu+(k/|k|)d_{|k|}\right]^2+
\left[\left(\Gamma+\delta\Gamma_{|k|}\right)/2\right]^2}+B,
\end{equation}
where $A$ was the amplitude of the central peak, $\beta_k$ was the
height of the $k\rm^{th}$ peak relative to the central peak,
$\Gamma$ was the width of the central peak, $\delta\Gamma_{|k|}$ was
the increase in width of the $l=0,2$ and $l=1,3$ overlapping peaks,
$d_{|k|}$ was the frequency spacing between the central peak and the
outer peaks and $B$ was a background term. Here $|k|=0$ represented
the central peak, $|k|=1$ represented the $l=0,2$ overlap, $|k|=2$
represented the diurnal sidebands and $|k|=3$ represented the
$l=1,3$ overlap. The relative height, $\beta_k$ was fixed at unity
for $k=0$ i.e. for the central peak. The increase in width,
$\delta\Gamma_{|k|}$ was fixed at $0$\,\textmu Hz when $|k|=0$ and 2
i.e. for the central peak and the diurnal sidebands. Finally $d_{0}$
was fixed at $0$\,\textmu Hz.

Following \citet{Chaplin2007} we assumed that the widths of the
diurnal sidebands were the same as the central peak and fixed their
frequencies at $\pm11.57$\,\textmu Hz. The Lorentzians for the
overlapping pairs have widths that are wider than the central peak
because of the influence of different components of a multiplet. For
example an $l=0$ mode, which has one component only, will overlap
with the 3 visible (in Sun-as-a-star data) $m$ components of an
$l=2$ mode at 3 different frequencies. The distances of the
overlapping $l=0,2$ and $l=1,3$ pairs from the central peak are
known as the small separations, $d_{l, l+2}$. We have used initial
guess values for the $l=0,2$ pair of $d_{0,2}=8.9$\,\textmu Hz and
for the $l=1,3$ pair of $d_{1,3}=15.9$\,\textmu Hz, based on the
average observed spacings over the frequency range of interest. For
the remainder of this paper we refer to this method of determining
the large separations as the ``autocorrelation method''.

The underlying profile of a mode in a frequency-power spectrum is an
asymmetric Lorentzian. Two important parameters in characterizing
each mode Lorentzian are the height and width. We then define the
power of the mode as the total area under the Lorentzian. The
autocorrelation function is most influenced by the modes with the
largest heights i.e. those modes around $3100$\,\textmu Hz. In
effect the obtained frequency separation represents a weighted
average over the frequency range of interest.

Although the statistics of a frequency-power spectrum is not
Gaussian the autocorrelation function is the sum of many points and
so the central limit theorem can be applied. Therefore, a standard
least squares fitting can be performed. However, the points in the
autocorrelation are highly correlated and, although this does not
bias the fitted parameters, the formal Hessian uncertainties are an
underestimate of the true uncertainties. Monte Carlo simulations,
similar to those described in Section \ref{section[method fit
frequencies]}, were performed to estimate a scaling factor by which
the Hessian uncertainties needed to be multiplied to better
represent the true errors of the fit.  The scaling factor showed no
systematic dependence on fill over the range examined here. We
therefore took the scaling factor to be the mean of the simulated
scaling factors. When calculating the autocorrelation function over
$2500\le \nu_{l,n} \le 3700$ the scaling factor was 3.5.

The middle panel of Fig. \ref{figure[examples]} shows an example of
the autocorrelation. A prominent peak is observed and the model
provides a good fit to the data. The fill of the plotted data is
relatively high for ground-based observations (0.88) and so overlaps
between the $l=0,2$ and $l=1,3$ modes are clearly visible, whereas
the daily harmonic peaks are suppressed.

\subsection{The PSPS method}\label{section[method PSPS]}

An alternative way of determining the large separations that is
frequently employed in asteroseismic studies
\citep[e.g.][]{Roxburgh2006, Roxburgh2009, Mosser2009, Hekker2010a,
Mathur2010, Mosser2010} is to calculate the power spectrum of the
power spectrum ($\rm PS\bigotimes PS$). A significant peak is
observed in the power spectrum of the power spectrum at half the
large separation. This method is equivalent to determining the
autocorrelation of the time series. \citet{Kholikov2008} used the
autocorrelation of time series to determine the acoustic radius,
$T$, which is related to the large separation, $\Delta\nu$, by the
following equation:
\begin{equation}
\Delta\nu=\frac{1}{2T}.
\end{equation}

The $\rm PS\bigotimes PS$ was determined, over the frequency range
of interest. We have oversampled the data by a factor of 10 and this
was done by adding zeroes to the end of the frequency range of
interest. A peak is clearly visible at half the large separation
(see the right-hand panel of Fig. \ref{figure[examples]}). A
Gaussian was fitted to this peak to determine the value of half the
separation. Two alternate ways of finding the central frequency of
the peak are by interpolation and by determining the centroid of the
peak. However, these methods were found to be more sensitive to the
background noise and so we have instead used the results obtained by
fitting a Gaussian. For the remainder of this paper we refer to this
method of determining $\Delta\nu$ as the ``power spectrum of the
power spectrum (PSPS) method''.

To determine whether the uncertainties from fitting a Gaussian to
the $0.5\Delta\nu$ peak are representative of the true uncertainties
Monte Carlo simulations were performed that were similar to those
described in Section \ref{section[method fit frequencies]}. The
simulations showed that the uncertainties were underestimated by a
factor that was dependent on the frequency range under
consideration. Different BiSON-like window functions were applied to
the simulated time series. As with the other two methods the scaling
factor was independent of the fill over the range of fills observed
here. When calculating the $\rm PS\bigotimes PS$ over $2500\le
\nu_{l,n} \le 3700$ the scaling factor was 1.8.

\section{Solar cycle variations in $\Delta\nu$}\label{section[results]}

\begin{figure}
\centering
  \includegraphics[width=0.45 \textwidth, clip]{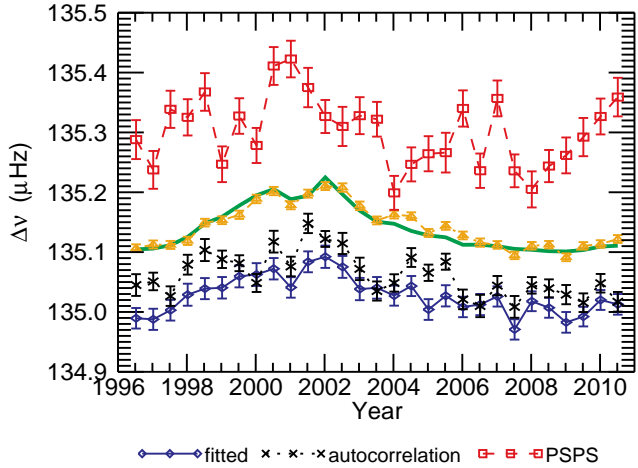}\\
  \caption{Large separations obtained
from the 185.5-d data sets using the fitted, autocorrelation, and
PSPS methods. For the fitted method the mean $\Delta\nu$ for modes
with $l\le 2$ was obtained, while the autocorrelation and PSPS
methods contain information from all modes visible in Sun-as-a-star
data. Scaled versions of the frequency shifts obtained by
\citet{Fletcher2010} and the 10.7-cm flux are plotted for comparison
purposes.}\label{large spacings}
\end{figure}

Fig. \ref{large spacings} shows the large separations obtained from
the 182.5-d data sets using the fitted, autocorrelation, and PSPS
methods. For the fitted method the mean $\Delta\nu$ for modes with
$l\le 2$ was obtained. As we have used Sun-as-a-star data the
autocorrelation and PSPS methods contain information from all
visible $l$. However, in practice, modes with $l\le2$ will dominate
as they are most prominent. Solar cycle variations are clearly
visible in the fitted and autocorrelation method results. The PSPS
method results are noisier and so it is difficult to see any solar
cycle variations.

There is an offset between the large separations obtained by the
three methods (also see the top panel of Fig. \ref{large spacings
flag}). The difference between the fitted and autocorrelation
methods can be explained in terms of the different frequency
dependence of the methods combined with the frequency dependence of
the large separations themselves. The autocorrelation method is
heavily weighted towards the most prominent modes. However, the
fitted method will be weighted towards lower-frequency modes since
the frequencies of these modes can be obtained more precisely than
high-frequency modes (because they have longer damping times and
consequently narrower widths in frequency-power spectra). Weighting
the linear fit between $n$ and $\nu_{n,l}$ by a combination of mode
height and the errors associated with the frequencies reduces the
disparity but only very marginally. The PSPS method is weighted by
the power (and not the height) of the modes. Mode power drops off
less rapidly with frequency than mode height and so the PSPS method
is weighted towards higher-frequency modes than the other two
methods. Since $\Delta\nu$ increases with frequency over the range
of frequencies considered here, the $\Delta\nu$ determined by the
PSPS method is larger than the $\Delta\nu$ obtained by the other
methods.

\begin{figure}
\centering
  \includegraphics[width=0.45\textwidth,
  clip]{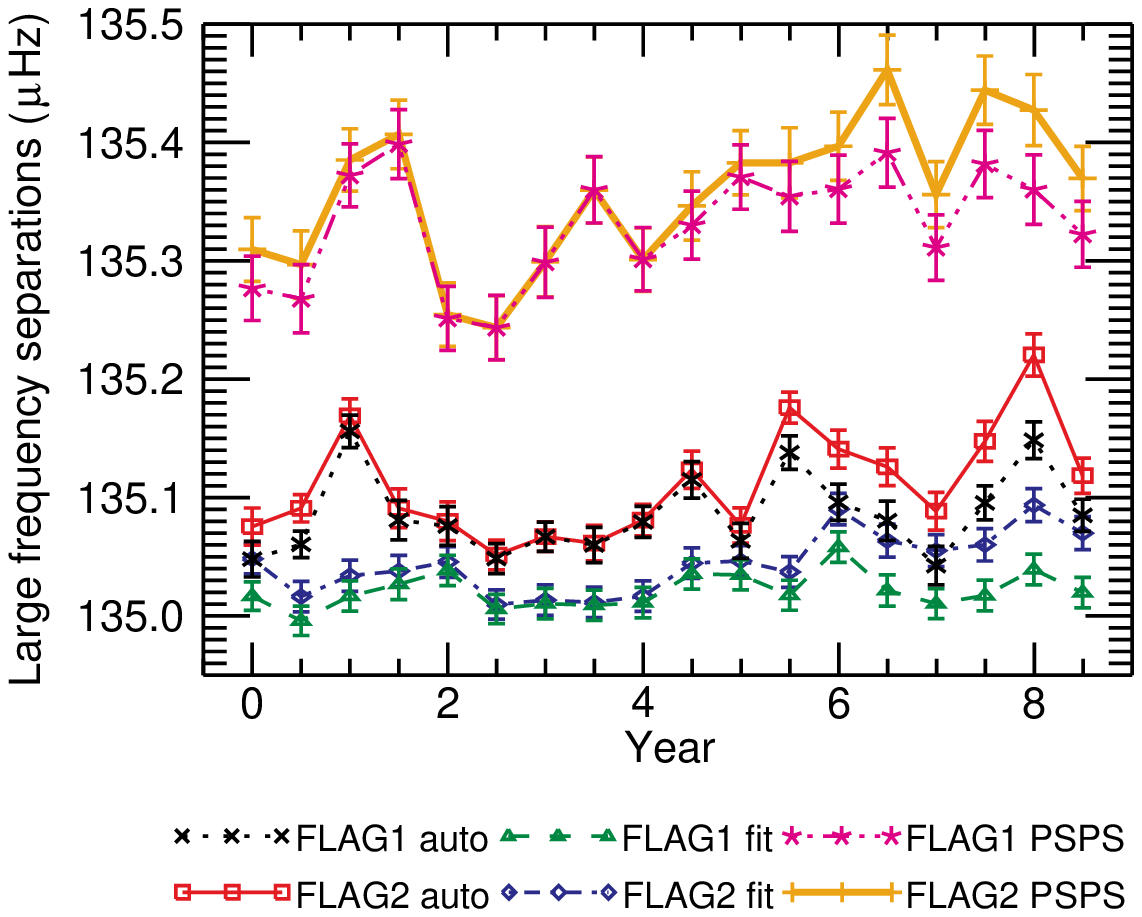}\\\vspace{0.5cm}
  \includegraphics[width=0.45\textwidth, clip]{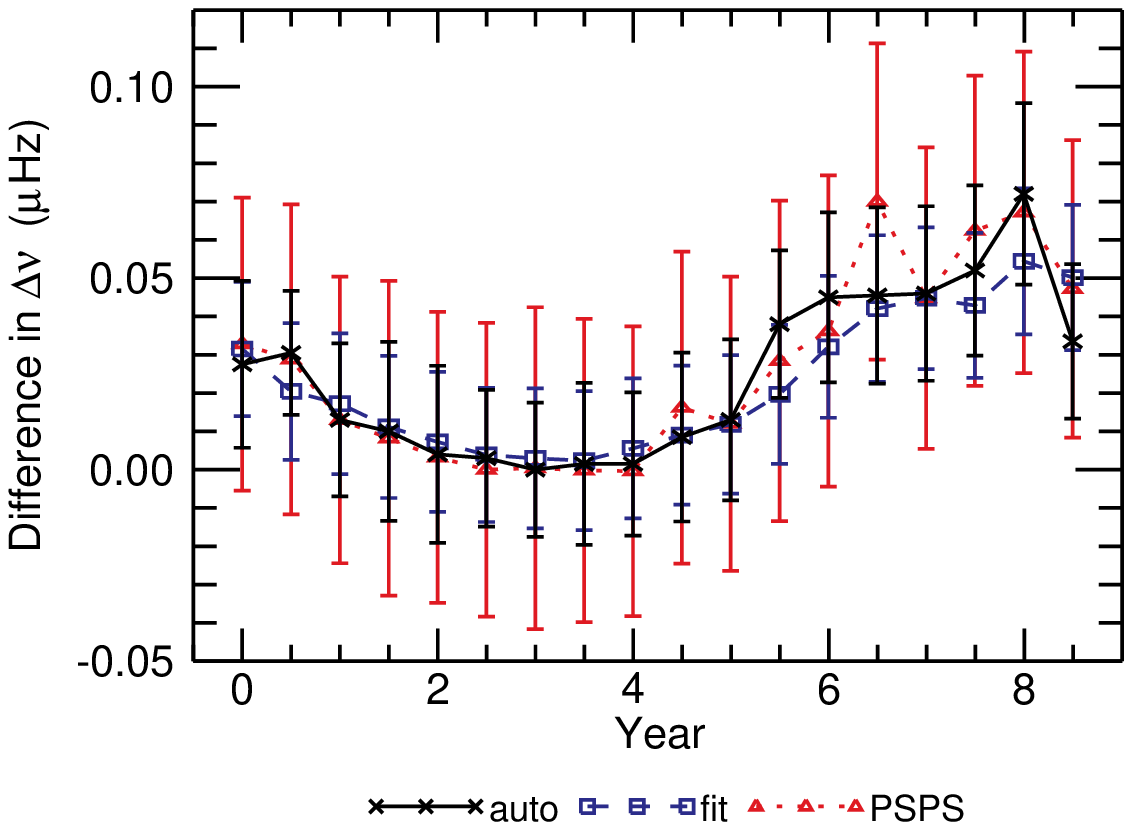}\\
  \caption{Top panel: Large separations observed
  in the FLAG1 and FLAG2 data. Bottom panel: Differences
  between the FLAG1 and FLAG2 large separations. For the fitted method the mean $\Delta\nu$
for modes with $l\le 2$ was obtained, while the autocorrelation and
PSPS methods contain information from all modes visible in
Sun-as-a-star data.
  }\label{large spacings flag}
\end{figure}

The large separations have also been obtained from two sets of data
that were simulated for the solar Fitting at Low Angular Degree
Group (FLAG) \citep[][top panel of Fig. \ref{large spacings
flag}]{Chaplin2006}. The data were simulated in the time domain. In
the first simulated data set (FLAG1) the properties of the
oscillations were constant but in the second data set (FLAG2) the
oscillation properties were modulated by the solar cycle. For the
fitted method the mean $\Delta\nu$ for modes with $l\le 2$ was
obtained while the autocorrelation and PSPS methods contain
information from all visible $l$. The large separations determined
by the autocorrelation and PSPS methods vary significantly more than
the large separations determined using the fitted method. This
indicates that the autocorrelation  and PSPS methods are more
sensitive to the realisation noise.

The bottom panel of Fig. \ref{large spacings flag} shows the
difference between the large separations obtained in the FLAG2 data
set and the large separations obtained from the FLAG1 data set. This
difference uncovers the simulated solar cycle effect, which alters
the large separations by about $0.05$\,\textmu Hz. It should be
remembered that this data set is only $\sim 9$\,yr in length and so
does not quite cover an entire solar cycle.

\subsection{Predicting variations in the large
separations using solar cycle frequency correction
techniques}\label{section[model]}

\citet{Broomhall2009a} corrected Sun-as-a-star p-mode frequencies
for the effect of the solar cycle. Here, we have reversed the
procedure applied to implement these corrections in order to predict
the frequencies of p modes in Sun-as-a-star data at a particular
level of activity. We have used the frequencies quoted in table 3 of
\citet{Broomhall2009} as a reference set of frequencies. These
frequencies were determined by fitting 23\,yr of BiSON data using
the methods described in \citet{Fletcher2009}. The raw fitted
frequencies were then corrected, using the methods described in
\citet{Broomhall2009a}, to give the frequencies that would have been
observed at the canonical quiet Sun activity level, and it is these
frequencies that we use here. The correction performed by
\citet{Broomhall2009a} comes in three parts: a linear solar cycle
correction, a ``devil-in-the-detail'' correction, and a
Sun-as-a-star correction.

The linear solar cycle correction is well known and can be used to
correct mode frequencies to a nominal activity level. The correction
is based on the assumption that variations in global activity
indices can be used as proxies for low-$l$ frequency shifts. It is
also assumed that the correction can be parameterised as a linear
function of the chosen activity measure. Here we have used the
10.7-cm flux \citep{Tapping1990} for which these assumptions are
robust at the level of precision of the data \citep[][and references
therein]{Broomhall2009a}. The canonical quiet-Sun value of the radio
flux is fixed from historical observations at $\langle
A(t)\rangle_c=64\times 10^{-22}\,\rm W\,m^{-2}\,Hz^{-1}$. The size
of the linear correction is also dependent on mode frequency and
$l$. Here we have used the frequency dependence derived by
\citet{Chaplin2001, Chaplin2004}. The $l$ dependence is small for
$l\le3$ but is included nonetheless and occurs because of
differences in the mode inertia \citep{JCD1991}.

The ``devil-in-the-detail'' correction accounts for cross-talk
between the variations of different mode parameters over the solar
cycle, and the distribution of activity levels over the period of
observations. As we are using relatively short time series here,
this effect will be negligible and so this correction was not
performed.

The Sun-as-a-star correction accounts for the fact that when using
Sun-as-a-star observations the rotationally split components of a
mode have different visibilities. As the plane of the Sun's rotation
axis is nearly perpendicular to the line-of-sight, only modes where
$l+m$ is even have significant visibility. Furthermore, estimates of
the centroid frequencies are dominated by the $|m|=l$ components, as
they are most prominent. The difference between the centroid and
fitted frequencies depends on the level of activity over the period
the observations were made. When the solar activity is at a minimum
the components are observed to be in a near-symmetrical arrangement
and so the fitted centroid is close to the true centroid frequency.
However, at moderate to high activity levels this is not the case as
the mode components are not arranged symmetrically. For example, the
$|m|=2$ components of an $l=2$ mode will experience a larger shift
at high activities than the $m=0$ component. The magnitude of the
observed asymmetry is related to the inhomogeneous distribution of
the magnetic activity over the solar surface and the spherical
harmonic associated with each visible $m$ component. Therefore, the
fitted frequencies differ from the true centroids by an amount that
is dependent on $l$. \citet{Appourchaux2007} describe how to make a
Sun-as-a-star correction, which is determined using the so-called
$a$ coefficients that are found from fits for an unresolved Sun.

\begin{figure}
\centering
  \includegraphics[width=0.45\textwidth, clip]{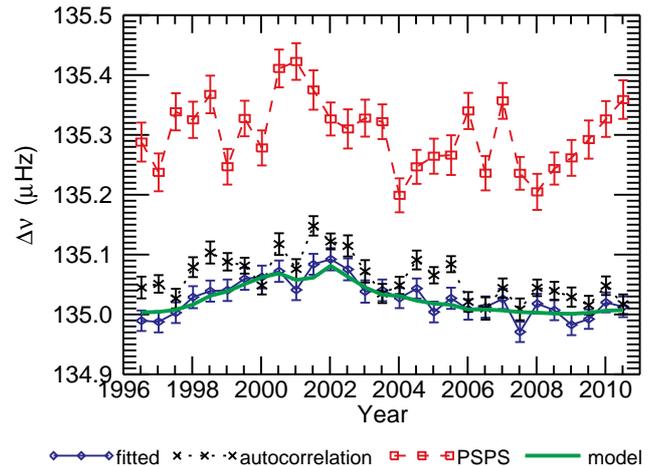}\\
  \caption{Comparison between the predicted and observed $\Delta\nu$. The
  different symbols represent the results of different methods (see
  legend) and the thick solid green line represents the model.
  For the fitted method the mean $\Delta\nu$
for modes with $l\le 2$ was obtained, while the autocorrelation and
PSPS methods contain information from all modes visible in
Sun-as-a-star data.
  }\label{figure[predictions]}
\end{figure}

Once the frequencies for a particular time series at a known
activity level have been predicted we can use these frequencies to
determine the average large frequency separation using the fitted
method. The predicted large separations are plotted in Fig.
\ref{figure[predictions]} along with the observed $\Delta\nu$
determined by the three methods. For the fitted method the mean
$\Delta\nu$ for modes with $l\le 2$ was obtained, while the
autocorrelation and PSPS methods contain information from all modes
visible in Sun-as-a-star data. The agreement between the predicted
large separations and the large separations observed using the
fitted method is very good. The good agreement between the predicted
and observed large separations implies that the observed variation
can be explained in terms of the observed shifts of the individual
mode frequencies.

Notice that the observed large separations show more short-term
variability than the predicted $\Delta\nu$ in the declining phase of
cycle 23 and the minimum between cycles 23 and 24. It has long been
observed that p-mode frequencies respond to short-term changes in
activity \citep[e.g][]{Rhodes2002}. Furthermore, a similar behaviour
was observed in the solar-cycle frequency shifts by
\citet{Broomhall2009} and \citet{Fletcher2010}. Fig. \ref{large
spacings} shows that the quasi-biennial variations observed in the
large separations are in agreement with those observed in the
frequency shifts by \citeauthor{Broomhall2009} and
\citeauthor{Fletcher2010}. This implies that the shorter-term
variations show some frequency dependence. However, the amplitude of
the shorter-term variations in $\Delta\nu$ is smaller than the
amplitude of the 11-yr variation, indicating that the frequency
dependence of the shorter-term variation is weaker than the
frequency dependence of the 11-yr cycle. This is in agreement with
the results of \citet{Fletcher2010}.

\begin{figure}
\centering
  \includegraphics[width=0.45\textwidth, clip]{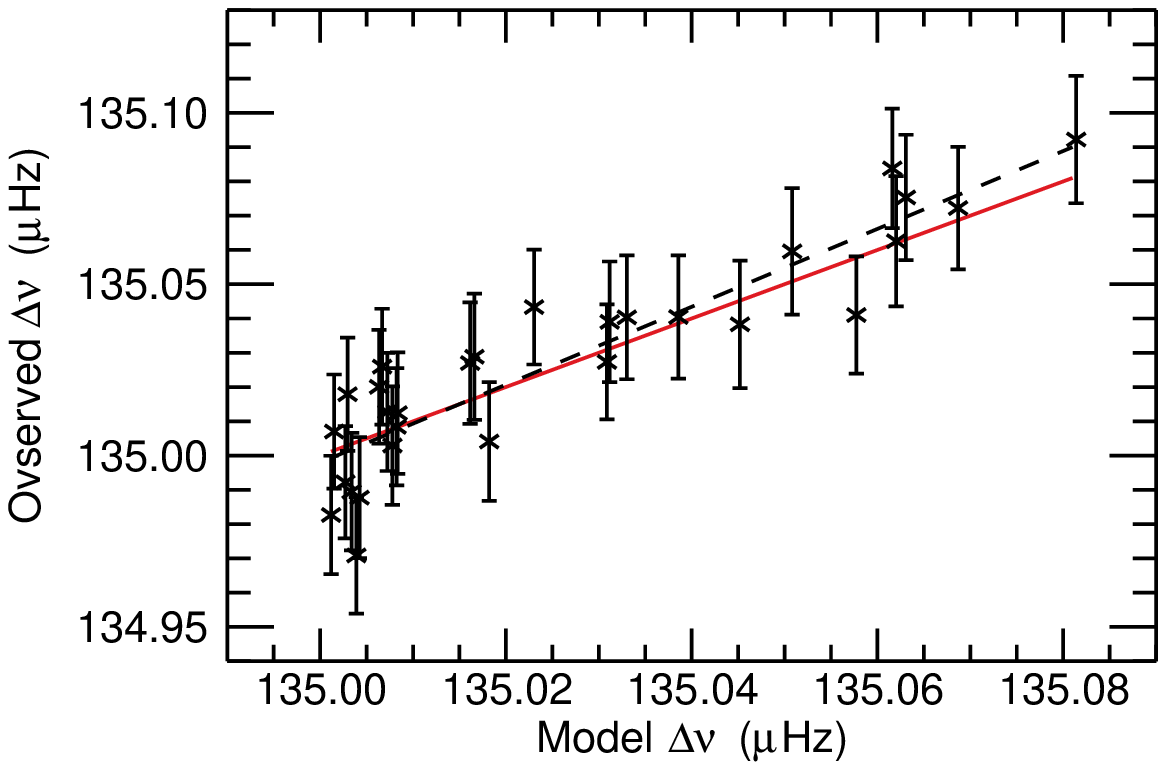}\\
  \caption{Comparison between the modelled and fitted method observed $\Delta\nu$ obtained
  over $2500\le\nu_{l,n}\le3700$\,\textmu Hz using modes with $l\le2$.
  The symbols represent the observed $\Delta\nu$ and the dashed line represents the linear fits between the
  model and observed $\Delta\nu$. The 1:1 relation has been also
  been plotted (the red solid line) to guide the eye.}\label{figure[model vs obs]}
\end{figure}

Fig. \ref{figure[model vs obs]} shows a direct comparison between
the model and fitted method observed $\Delta\nu$ using modes with
$l\le2$. Although not shown here a similar analysis was performed
for the autocorrelation and PSPS methods. Linear fits between the
modelled and observed $\Delta\nu$ were performed and the gradients
of the linear fits were within $1\sigma$ of unity for all three
methods. This suggests that the magnitude of the solar cycle effect
is the same for each method of deriving $\Delta\nu$ and that all
three methods produce variations in $\Delta\nu$ that are in
agreement with the model, even if the absolute values of $\Delta\nu$
differ (e.g. Fig. \ref{figure[predictions]}).

\subsection{The $l$ dependence of the solar cycle variations in the large
separations}\label{section[l dependence]}

\begin{figure}
\centering
  \includegraphics[width=0.45\textwidth, clip]{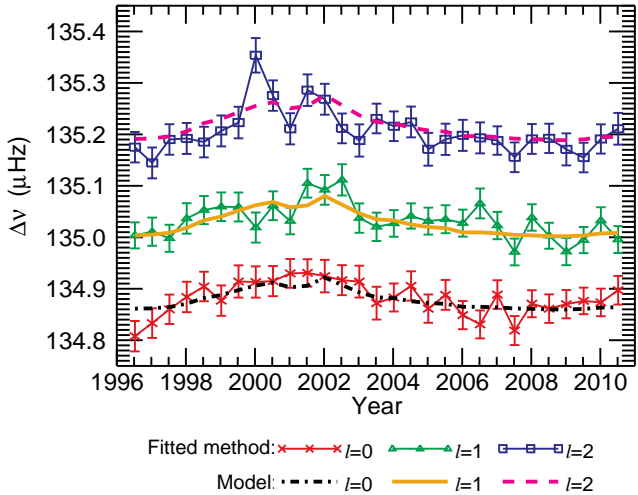}\\
  \caption{Large separations observed for individual $l$. The
  results of the fitted method are plotted for each $l\le2$, as are
  the model predictions (see legend).}\label{figure[l dependence]}
\end{figure}

As we are using Sun-as-a-star data the large separations cannot be
determined for each $l$ separately using the autocorrelation and
PSPS methods. However, the fitted method can be used to examine the
$l$ dependence of the variation in $\Delta\nu$. Fig. \ref{figure[l
dependence]} shows the solar cycle variations in the large
separations for individual $l$. There is an offset between the
different large separations observed in the different $l$,
demonstrating the $l$ dependence of the large separations.
$\Delta\nu$ increases with increasing $l$ because the depth of the
cavity in which the modes are trapped decreases and so the acoustic
radius, which is inversely proportional to $\Delta\nu$, also
decreases with increasing $l$.

\begin{figure*}
\centering
  \includegraphics[width=0.3\textwidth, clip]{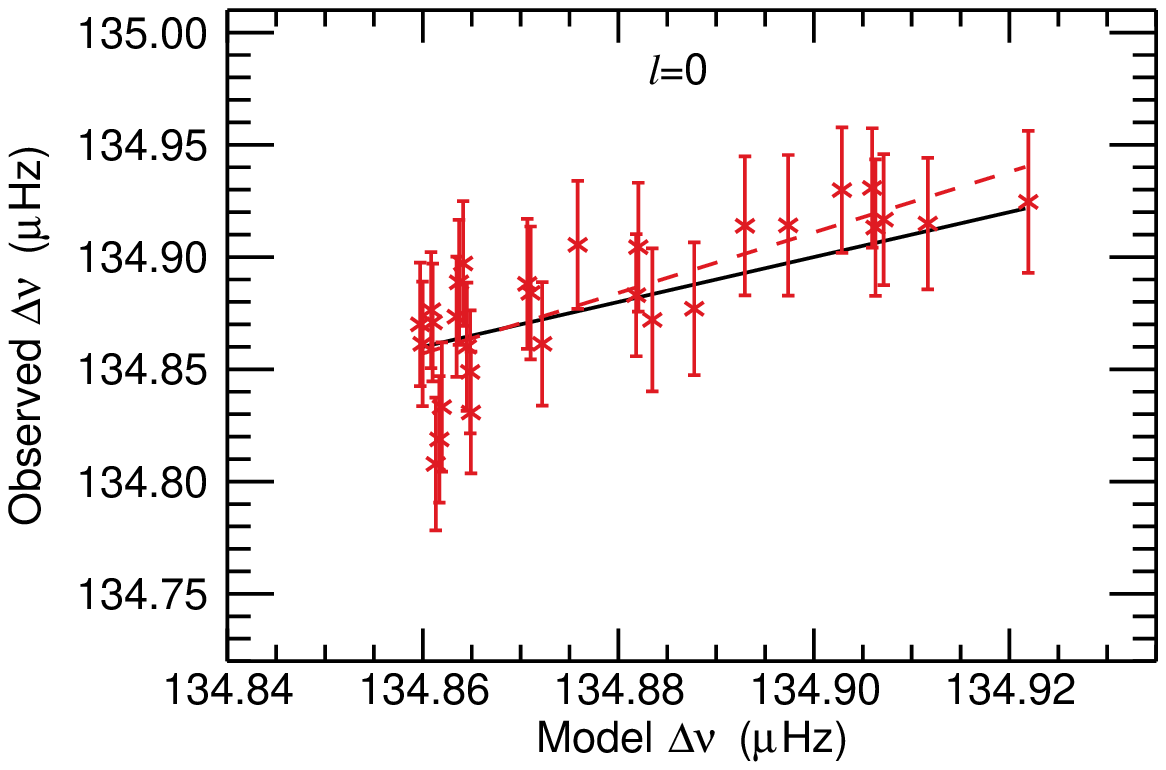}
  \includegraphics[width=0.3\textwidth, clip]{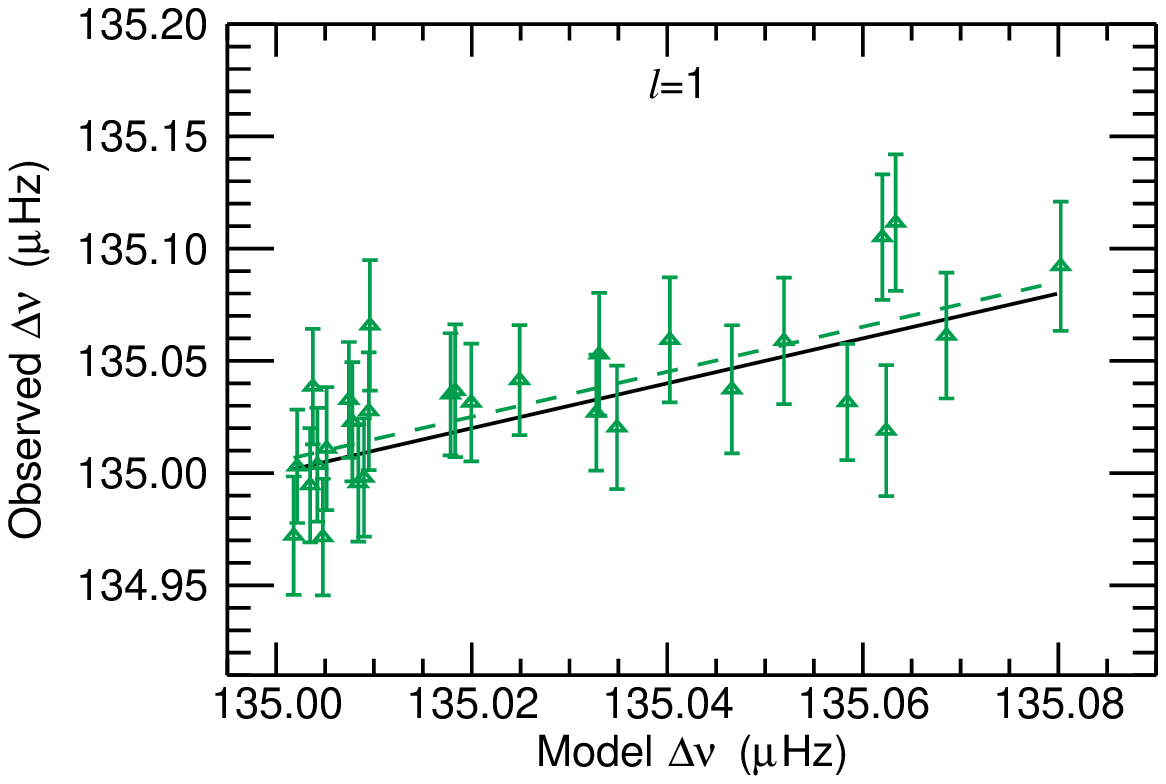}
  \includegraphics[width=0.3\textwidth,
  clip]{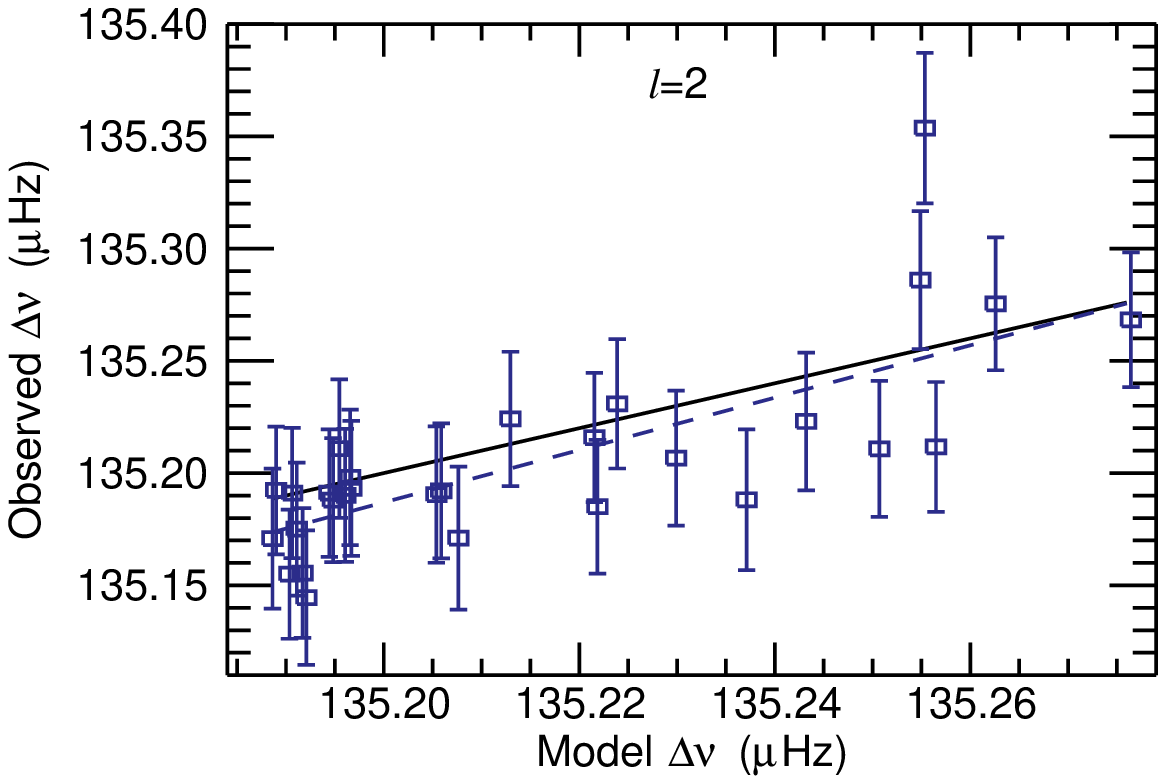}\\
  \caption{Comparison between the modelled and observed $\Delta\nu$ obtained using the fitted method.
  The results for the different $l$ are plotted in separate panels.
  In each panel the symbols represent the observed $\Delta\nu$ and
  the dashed lines represent a linear fit between the modelled and
  observed $\Delta\nu$. The 1:1 relation has been also
  been plotted in each panel (the black solid lines)
  to guide the eye.}\label{figure[model vs obs l]}
\end{figure*}

The individual-$l$ large separations are generally well reproduced
by the model described in Section \ref{section[model]}. Figs
\ref{figure[l dependence]} and \ref{figure[model vs obs l]} show
that the magnitude of the observed 11-yr cycle variations increase
slightly with $l$. For the model values the $l=0$ peak-to-trough
variation is $0.06$\,\textmu Hz, whereas the $l=1$ variation is
$0.08$\,\textmu Hz and the $l=2$ variation is $0.09$\,\textmu Hz.
Solar cycle shifts in mode frequencies are dependent on $l$ because
the size of the perturbation is inversely related to the mode
inertia, which decreases with increasing $l$
\citep[e.g.][]{Libbrecht1990}. However, this effect is negligible
over the range of $l$ considered here ($l\le2$). As mentioned in
Section \ref{section[model]} there is an $l$ dependence in the solar
cycle frequency shifts that occurs because magnetic activity is
inhomogenously spread across the solar surface. It is this $l$
dependence that causes the differences with $l$ in the amplitudes of
the solar cycle variations in $\Delta\nu$.

Fig. \ref{figure[model vs obs l]} shows a direct comparison between
the model and observed $\Delta\nu$ of each $l$. The model has been
computed separately for each $l$. Although offset from each other
the ranges of the axes are the same and so it is again possible to
see that the $l=2$ modes show more variation throughout the solar
cycle than the $l=1$, and, in particular, the $l=0$ modes. Linear
fits between the modelled and observed $\Delta\nu$ were again
performed and the gradients of the linear fits lie within $2\sigma$
of unity for all three $l$, with the $l=1$ modes showing the best
agreement.

\section{Impact of observational choices}\label{section[comparison]}

\citet{Kholikov2008} observed solar cycle variations in the acoustic
radius, which is inversely proportional to $\Delta\nu$, in Global
Oscillations Network Group (GONG) and Michelson Doppler Imager (MDI)
data. To allow a comparison with the results of \citet{Kholikov2008}
we have repeated the above analysis using the same frequency range
i.e. $2300\le\nu\le4300$\,\textmu Hz. \citeauthor{Kholikov2008}
determine the acoustic radius for individual $l$. Therefore we have
compared the \citeauthor{Kholikov2008} $l=0$ results with the $l=0$
$\Delta\nu$ found here. Before that we first discuss the impact of
changing the frequency range on the $l$-averaged results.

Fig. \ref{figure[separations alt]} shows the large separations found
using the autocorrelation, fitted, and PSPS methods, and the model
predictions. The fitted method was used to determine the average
$\Delta\nu$ for modes with $l\le2$. The change in frequency range
has increased the offset between the results of the three methods.
Changing the frequency range has had very little effect on the
$\Delta\nu$ determined using the autocorrelation method because the
autocorrelation method is weighted most heavily towards the most
prominent modes, at $\sim3100$\,\textmu Hz. However, the fitted
method large separations have been shifted downwards by
approximately $0.2$\,\textmu Hz and the short term structure of the
large separations has changed. The shift in the fitted method large
separations arises because the obtained $\Delta\nu$ are weighted
towards the lowest frequency modes in the range under consideration
and modes at $2300$\,\textmu Hz have a lower $\Delta\nu$ than modes
at $2500$\,\textmu Hz. The change in the short-term structure of the
fitted method large separations could be due to extra noise, which
is introduced into the results by including the higher frequency
modes, whose frequencies cannot be determined precisely because of
short mode lifetimes. It could also be because low-frequency modes
are more sensitive to realization noise, due to long mode lifetimes.
Another possibility is that the changes in the short-term structure
could reflect the frequency dependence of the shorter-term signal.

The large separations determined by the PSPS method have been
shifted upwards by approximately $0.12$\,\textmu Hz. As mentioned
earlier the PSPS method is weighted by mode power and the power of
modes in the range $3700-4300$\,\textmu Hz is, on average, higher
than the power of modes in the range $2300-2500$\,\textmu Hz. Since
modes in the frequency range $3700-4300$\,\textmu Hz have larger
$\Delta\nu$ than lower-frequency modes the large separations
determined by the PSPS method are increased.

\begin{figure}
\centering
  \includegraphics[width=0.45\textwidth, clip]{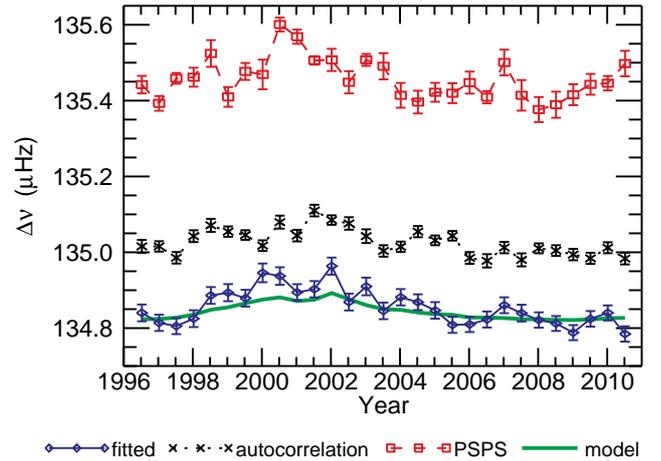}\\
  \caption{$\Delta\nu$ observed over a frequency range of $2300-4300$\,\textmu Hz. The different symbols represent the results of the different
  methods (see legend). Model predictions are also plotted.
  For the fitted
method the mean $\Delta\nu$ for modes with $l\le 2$ was obtained,
while the autocorrelation and PSPS methods contain information from
all modes visible in Sun-as-a-star data.}\label{figure[separations
alt]}
\end{figure}

Also plotted in Fig. \ref{figure[separations alt]} are the modelled
$\Delta\nu$ determined over the frequency range
$2300\le\nu_{l,n}\le4300$\,\textmu Hz and for modes with $l\le2$. As
expected they maintain approximate agreement with the fitted method
results (as opposed to the autocorrelation and PSPS results).
However, the agreement between the modelled and fitted $\Delta\nu$
is clearly not as good over this frequency range as was found in
Fig. \ref{figure[l dependence]} (over
$2500\le\nu_{l,n}\le3700$\,\textmu Hz).

\begin{figure}
\centering
  \includegraphics[width=0.45\textwidth, clip]{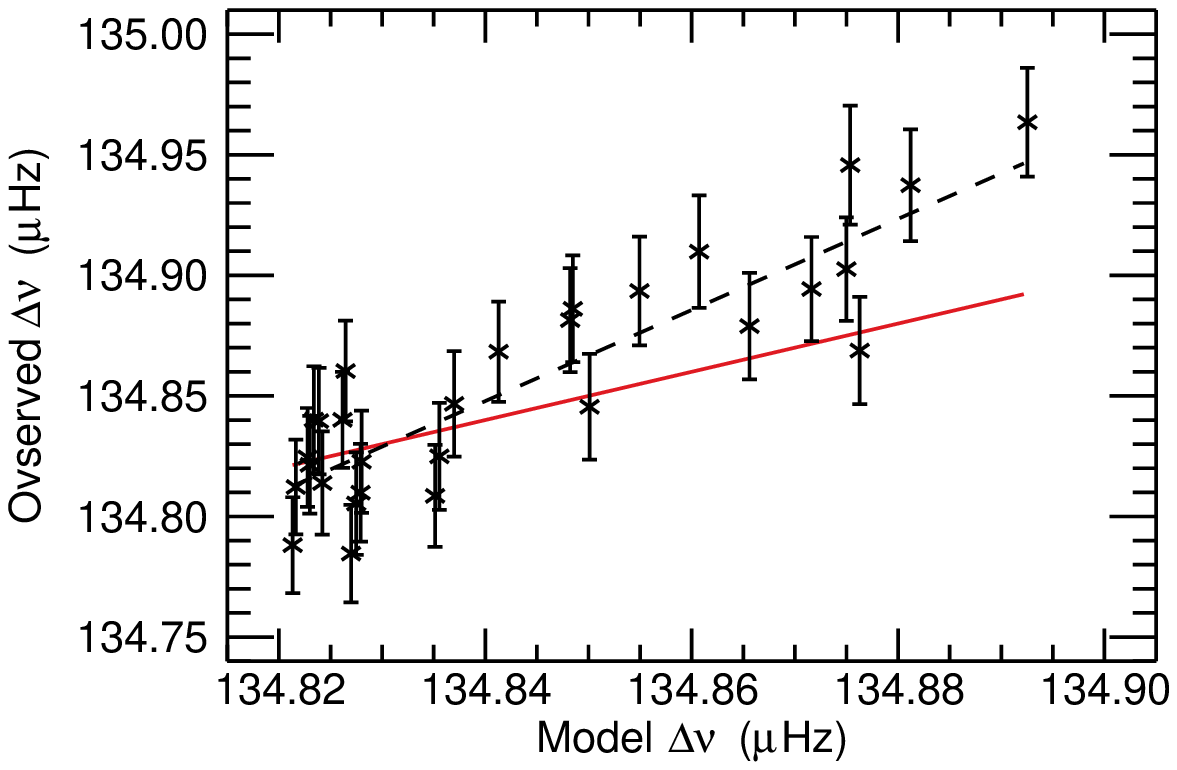}\\
  \caption{Comparison between the modelled and fitted method observed $\Delta\nu$ obtained
  over $2300\le\nu_{l,n}\le4300$\,\textmu Hz and $l\le2$.
  The symbols represent the observed $\Delta\nu$ and the dashed line represents the linear fits between the
  model and observed $\Delta\nu$. The 1:1 relation has been also
  been plotted (the red solid line) to guide the eye.}\label{figure[model vs obs alt]}
\end{figure}

Fig. \ref{figure[model vs obs alt]} shows a direct comparison
between the model and fitted method observed $\Delta\nu$ for
$2300\le\nu_{l,n}\le4300$\,\textmu Hz and $l\le2$. Once again,
although not plotted here a similar analysis was performed for the
autocorrelation and PSPS methods. The gradients of the linear fits
all show departures from unity: the gradient of the fitted method is
$4.7\sigma$ from unity, the autocorrelation is $1.6\sigma$ from
unity, and the PSPS method is $4.5\sigma$ from unity. The observed
$\Delta\nu$ show more variation than was predicted by the model.

Further investigation reveals that this is because the lower limit
of the frequency range has been changed from $2500$\,\textmu Hz to
$2300$\,\textmu Hz, which implies that the frequency dependence of
the solar cycle perturbation on mode frequencies does not reflect
the perturbation observed in BiSON data at low frequencies.

\begin{figure}
\centering
  \includegraphics[width=0.45\textwidth, clip]{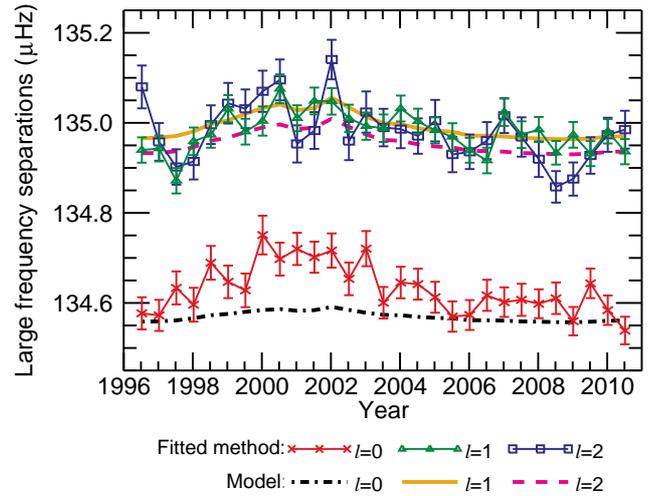}\\
  \caption{$\Delta\nu$ observed for individual $l$ over
  $2300-4300$\,\textmu Hz. Model predictions are also plotted.}\label{figure[l dependence alt]}
\end{figure}

Fig. \ref{figure[l dependence alt]} shows the $l$-dependence of the
large separations using the frequency range
$2300\le\nu_{n,l}\le4300$\,\textmu Hz. Comparison with Fig.
\ref{figure[l dependence]} shows that the $l=1$ large separations
are less sensitive to the change in frequency range than the $l=0$
and $l=2$ modes. Notice that the models do not represent the
observed large separations as well over this frequency range.

\begin{figure*}
\centering
  \includegraphics[width=0.3\textwidth, clip]{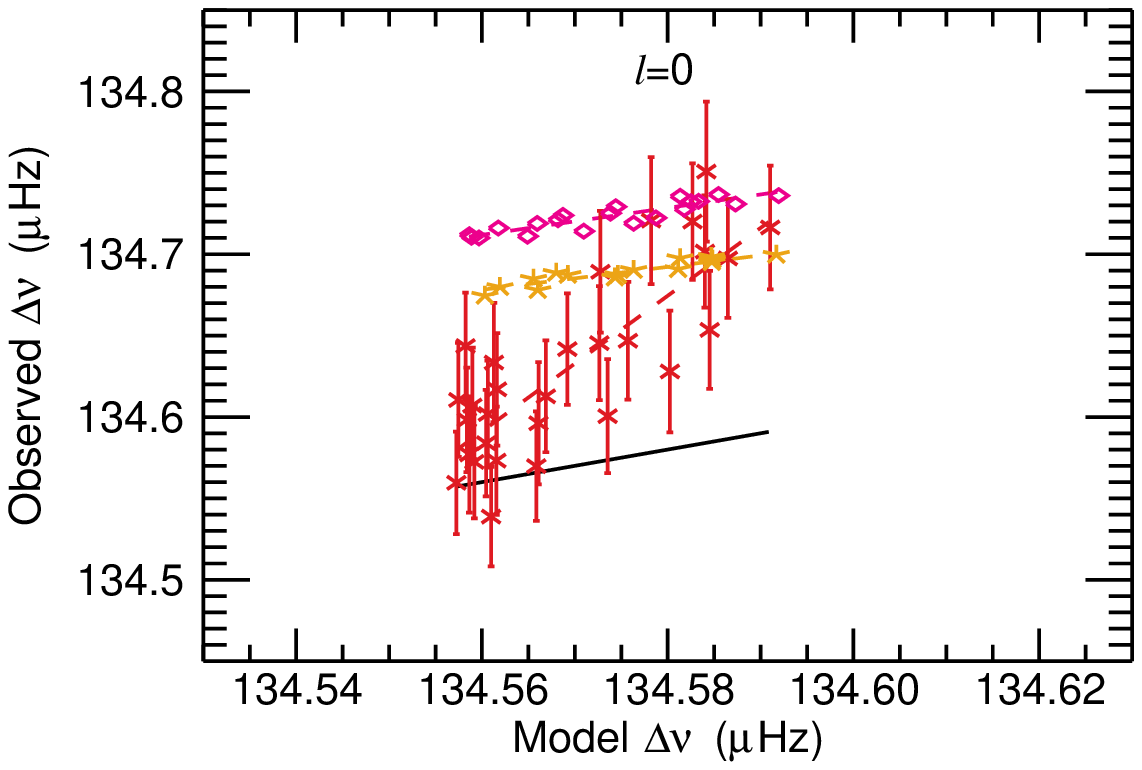}
  \includegraphics[width=0.3\textwidth, clip]{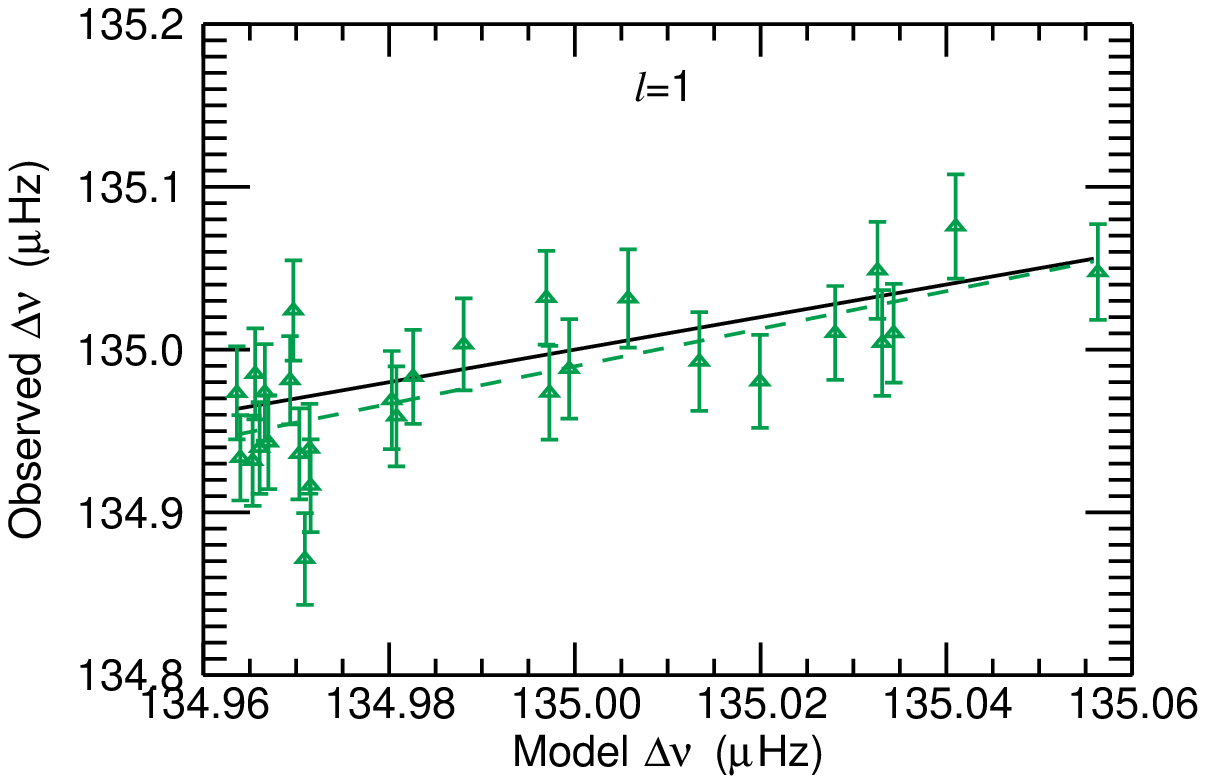}
  \includegraphics[width=0.3\textwidth,
  clip]{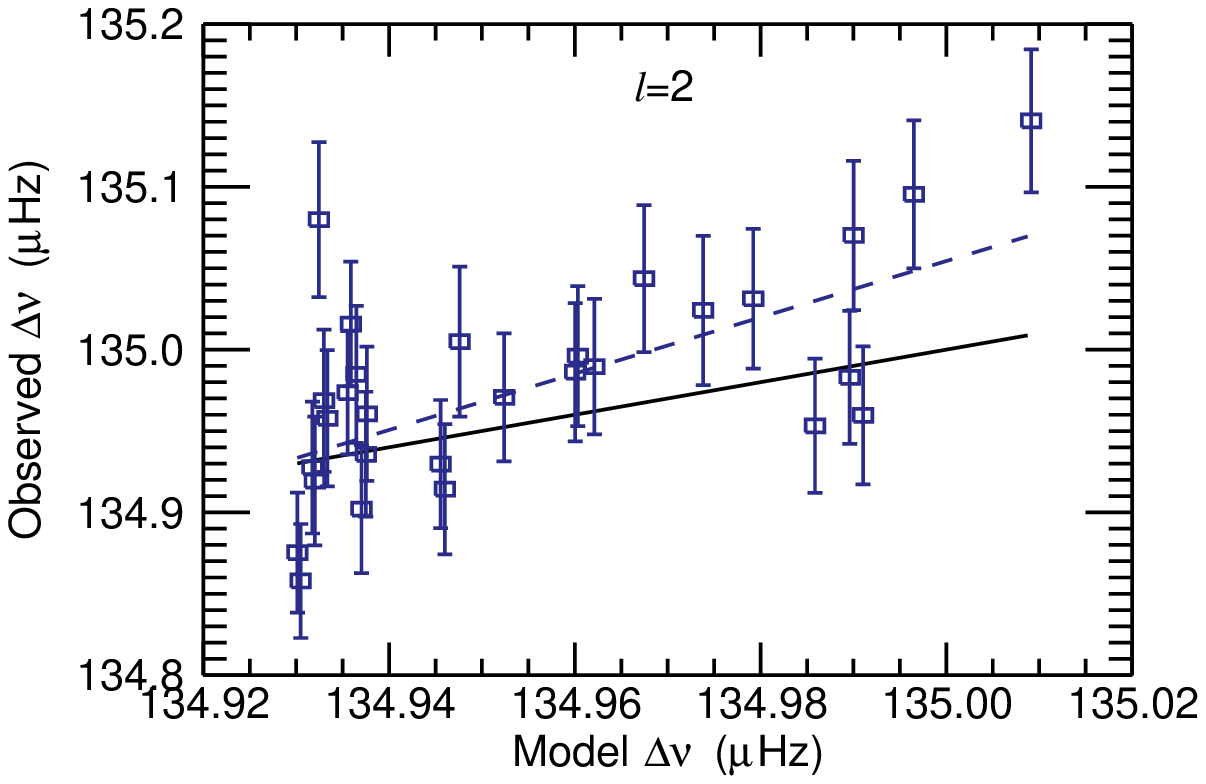}\\
  \caption{Comparison between the modelled and observed $\Delta\nu$
  obtained using the fitted method over $2300\le\nu_{l,n}\le4300$\,\textmu Hz.
  The results for the different $l$ are plotted in separate panels.
  In each panel the symbols represent the observed $\Delta\nu$ and
  the dashed lines represent a linear fit between the modelled and
  observed $\Delta\nu$. The 1:1 relation has been also
  been plotted in each panel (the black solid lines)
  to guide the eye. In the $l=0$ panel the red crosses represent the
  results obtained using the BiSON data. Also plotted are
  smoothed versions of the
  results obtained by \citet{Kholikov2008} for the GONG (pink
  diamonds) and MDI (orange asterisks) data.}\label{figure[model vs obs l alt]}
\end{figure*}

Fig. \ref{figure[model vs obs l alt]} shows a direct comparison
between the model and observed $\Delta\nu$ for each $l$. Although
the gradient of the linear fit to the $l=1$ mode results lies within
$1\sigma$ of unity the gradients of the fits to the $l=0$ and $l=2$
modes differ significantly from unity: $4.9\sigma$ for $l=0$ and
$2.2\sigma$ for $l=2$. Again the observed $\Delta\nu$ show more
variation than was predicted by the model. Therefore, the model does
not represent the data for $l=0$ and $l=2$ modes over this frequency
range.

\citet{Kholikov2008} use the autocorrelation of time series to
obtain the acoustic radius. This method is analogous to the PSPS
method used here. \citeauthor{Kholikov2008} use resolved Sun GONG
and MDI data and so are able to isolate individual $l$. Here, since
we are using Sun-as-a-star data we can only determine $\Delta\nu$
for individual $l$ using the fitted method. Therefore, in the
left-hand panel of Fig. \ref{figure[model vs obs l alt]} we compare
the $l=0$ results of \citet{Kholikov2008} with the $l=0$ $\Delta\nu$
determined using the fitted method. \citeauthor{Kholikov2008}
determined the offset for 36\,d (GONG) and 72\,d (MDI) time series.
To allow a comparison with the results derived here we have plotted
the weighted mean of the results of \citeauthor{Kholikov2008} over 5
and 3 points respectively, where the weights were given by the
errors of the observations. The range of $\Delta\nu$ observed in the
BiSON data is significantly larger than is observed by
\citeauthor{Kholikov2008}. A linear fit between the results of
\citeauthor{Kholikov2008} and the model was performed. The gradients
of both fits were significantly shallower than unity: $17.6\sigma$
away from unity for the GONG results and $13.8\sigma$ for the MDI
results. It should be noted that the uncertainties associated with
the results of \citeauthor{Kholikov2008} were very small. While, to
the eye, the gradients of the slopes look similar to unity, relative
to the size of the uncertainties, the departures from unity are
significant.

The GONG and MDI $\Delta\nu$ found by \citeauthor{Kholikov2008} are
offset from the BiSON results obtained here and from each other.
This suggests that not only do different methods of determining
$\Delta\nu$ result in different values (see e.g. Fig.
\ref{figure[model vs obs alt]}) but that also different data result
in different values of $\Delta\nu$. \citeauthor{Kholikov2008} were
unable to explain this offset. The autocorrelation method was used
to determine $\Delta\nu$ for 182.5-d sets of data observed by the
Global Oscillations at Low Frequencies \citep[GOLF;][]{Gabriel1995,
Jimenez2003, Garcia2005} instrument onboard the \emph{SOlar and
Heliospheric Observatory} (\emph{SOHO}) spacecraft. GOLF has been
collecting data since 1996 and so we have been able to analyze data
covering almost the entirety of solar-cycle 23, i.e., from 1996
April 11 to 2009 April 7. Although not shown here the BiSON and GOLF
results were in very good agreement and the differences between the
BiSON and GOLF $\Delta\nu$ were, on average, significantly smaller
than the associated error bars. It is possible, therefore, that the
main offset between the BiSON $\Delta\nu$ and the $\Delta\nu$
observed by \citeauthor{Kholikov2008} occurs because of the
different methodologies employed.

\section{Summary and Discussion}\label{section[discussion]}

The three different methods resulted in different values of
$\Delta\nu$. The results of \citet{Kholikov2008} were offset from
each other and from the BiSON data. The different values for the
different methods can be understood in terms of the weighting with
frequency of the methods and the frequency dependence of
$\Delta\nu$. However, this does not explain the difference between
the two sets of \citeauthor{Kholikov2008} results

The observed offsets could be important for asteroseismic studies,
which use the determined $\Delta\nu$ to infer stellar properties
such as radius and mass. \citet{Stello2009b} showed that the
following scaling relation holds over most of the HR diagram and
results in errors that are probably below 1 per cent,
\begin{equation}\label{equation[scaling relation]}
    \frac{\Delta\nu}{\Delta\nu_{\odot}}=\sqrt{\frac{M/M_{\odot}}{(R/R_{\odot})^3}},
\end{equation}
where $M$ is the mass of a star, $R$ is the radius of as star and
$\odot$ denotes the values for the Sun. The farthest separated
like-for-like values considered here are the fitted method and PSPS
method results for the frequency range
$2300\le\nu_{n,l}\le4300$\,\textmu Hz, which are on average
separated by $0.66$\,\textmu Hz. If we take the fitted method
results to represent those of the Sun and the PSPS method results to
represent those of another star and assume that the $M/M_\odot=1$,
equation \ref{equation[scaling relation]} implies that the ``other''
star is 0.3\,per cent larger than the Sun, which is within the
errors implied by \citet{Stello2009b}.

The fitting method appears to be the cleanest way of determining
$\Delta\nu$ as it is less sensitive to the noise realization than
the other two methods. However, it is not always possible to
determine the frequencies of the individual modes because of poor
signal-to-noise levels and poor fill. If this is the case, for the
data examined here, the autocorrelation method produced more stable
results than the PSPS method.

Solar cycle changes in $\Delta\nu$ were visible in the results of
the fitted and autocorrelation methods. This again suggests that
care is needed in using $\Delta\nu$ in determining properties of
stars. The amplitude of the solar cycle effect is smaller than
$0.66$\,\textmu Hz (see Fig. \ref{large spacings}) and, therefore,
using the same logic as above, the solar cycle variations in the Sun
would result in changes in stellar radii that are within error
estimates. However, the Sun is a relatively quiet star and so it is
possible that $\Delta\nu$ would vary more in other stars, resulting
in systematic errors in estimated masses and radii.

We have shown that the observed solar cycle changes in $\Delta\nu$
can be predicted using a simple model, which is valid over the
frequency range $2500\le\nu_{n,l}\le3700$\,\textmu Hz. This implies
that the changes in $\Delta\nu$ are due to the frequency dependence
of the solar cycle changes in $\nu_{l,n}$. The correlation with the
model is less good over the frequency range
$2300\le\nu_{l,n}\le4300$\,\textmu Hz. This is mainly because the
model underestimates the size of the perturbation in $\Delta\nu$ at
low frequencies, suggesting that the BiSON data do not follow the
assumed frequency dependence at low frequencies. This could be
because the assumed frequency dependence was derived using
intermediate- and high-degree GONG data and not (low-$l$) BiSON data
\citep{Chaplin2001, Chaplin2004}. The model also breaks down to
different extents for the different $l$. This could be because the
inertia scaling is not correct for the BiSON data at low
frequencies.

Recently \citet{Garcia2010} observed signatures of a stellar
activity cycle in asteroseismic data obtained by the Convection
Rotation and Planetary Transits \citep[CoRoT; e.g.][]{Michel2008}
space mission. With the prospect of longer asteroseismic data sets
($\sim3.5\,\rm yr$) becoming available through, for example, Kepler
\citep{Chaplin2010, Koch2010} there will be opportunities to observe
activity cycles in other stars. These observations will provide
constraints for models of stellar dynamos under conditions different
from those in the Sun.

\section*{Acknowledgements}

The authors thank S. Kholikov for providing the results of their
paper: \citet{Kholikov2008}. This paper utilizes data collected by
the Birmingham Solar-Oscillations Network (BiSON). We thank the
members of the BiSON team, both past and present, for their
technical and analytical support. We also thank P. Whitelock and P.
Fourie at SAAO, the Carnegie Institution of Washington, the
Australia Telescope National Facility (CSIRO), E.J. Rhodes (Mt.
Wilson, Californa) and members (past and present) of the IAC,
Tenderize. BiSON is funded by the Science and Technology Facilities
Council (STFC). The authors also acknowledge the financial support
of STFC.

\bibliographystyle{mn2e}
\bibliography{large_separations}
\end{document}